\documentclass[pra,longbibliography,showpacs,superscriptaddress,twocolumn,notitlepage, showkeys]{revtex4-1}
   
\usepackage{amsmath}
\usepackage{amsfonts}
\usepackage{graphicx, subfigure}
\usepackage{natbib} 
\usepackage{color}
\usepackage[usenames,dvipsnames]{xcolor}
\usepackage{mathtext}
\usepackage{wrapfig}
\usepackage{amsfonts,longtable}
\usepackage{amsmath}
\usepackage{amssymb}
\usepackage{bm}
\usepackage [basic,optics, gate, box]{circ}

\DeclareMathOperator{\df}{d\hspace{-0.8 mm}}
\DeclareMathOperator{\fb}{\textit{f}_{\mbox{\tiny B}}}
\DeclareMathOperator{\const}{const}
\DeclareMathOperator{\sinc}{sinc}

\DeclareMathOperator{\VH}{\mbox{$\widehat{V}$}\hspace{0.8 mm}} 
\DeclareMathOperator{\JH}{\mbox{$\widehat{J}$}\hspace{0.8 mm}} 
\DeclareMathOperator{\EH}{\mbox{$\widehat{\varepsilon}$}\hspace{0.8 mm}}

\begin{document}

\title{Adaptive Filtering to Enhance Noise Immunity of Impedance/Admittance Spectroscopy: Comparison with Fourier Transformation}
\author{Daniil D. Stupin}
\email{Stu87@ya.ru}
\email{Stupin@spbau.ru}
\affiliation{St. Petersburg Academic University, Khlopina 8/3, 194021 St. Petersburg, Russia}

\author{Sergei V. Koniakhin}
\affiliation{St. Petersburg Academic University, Khlopina 8/3, 194021 St. Petersburg, Russia}
\affiliation{Ioffe Physical-Technical Institute of the Russian Academy of Sciences, 194021 St.~Petersburg, Russia}

\author{Nikolay A. Verlov}
\affiliation{St. Petersburg Academic University, Khlopina 8/3, 194021 St. Petersburg, Russia}
\affiliation{Petersburg Nuclear Physics Institute NRC "Kurchatov Institute", Gatchina, St.\ Petersburg 188300, Russia}

\author{Michael V. Dubina}
\affiliation{St. Petersburg Academic University, Khlopina 8/3, 194021 St. Petersburg, Russia}
\affiliation{Peter the Great St. Petersburg Polytechnic University, Polytechnicheskaya str. 29, St. Petersburg, 195251, Russia}
\begin{abstract} 
The time domain technique for impedance spectroscopy consists in computing excitation voltage and current response Fourier images by fast or discrete Fourier transform and calculating their relation. Here we propose an alternative method for excitation voltage and current response processing for deriving system impedance spectrum based on fast and flexible adaptive filtering method. We show the equivalence between the problem of adaptive filter learning and deriving system impedance spectrum. To be specific we express the impedance via the adaptive filter weight coefficients. The noise canceling property of adaptive filtering has been also justified. Using the RLC circuit as a model system we experimentally show that adaptive filtering yields correct admittance spectra and elements ratings in the high noise conditions when Fourier transform technique fails. Providing the additional sensitivity for impedance spectroscopy, adaptive filtering can be applied to otherwise impossible to interpret time-domain impedance data. The advantages of adaptive filtering were justified with the practical living-cell impedance measurements.

\end{abstract}

\maketitle 
\section{Introduction}

\ Electrical  immittance (impedance and admittance) spectroscopy  \cite{Barsoukov,Lvovich} (EIS) is a powerful tool for electronic device diagnostics  \cite{Lebedev_Eng,Poklonski_Eng,Tunnel_IS}, materials characterization  \cite{Berman_Eng,QuantumDots, Ceramics,Ceramics2,Size_Effect,wavelet_cell}, electrochemistry \cite{barbero2007evidence,lelidis2005effect,CPE}, alternative energy sources investigation   \cite{Wang, SolarCell,tian2016enhanced,SolarCellBook}, and experiments in biophysics  \cite{Giaever2,Geaver,BioImp,rivnay2015using,weckstrom1992measurement,jin2016impedimetric}. 
The research of a dynamic systems have become a hot topic of impedance spectroscopy in recent years. The excitation voltage (EV)  with broad frequency spectrum like rectangular pulses, peaks, various types of noise, linear sweep or superposition of sine waves are commonly used for time resolved immittance spectroscopy. Employing such excitation signals allows scanning the sample in the wide frequency range, which gives significant advantage in performance and time resolution with respect to single-sine methods. To obtain the immittance spectrum (IS) from the collected data the  fast/discrete Fourier transform (FFT/DFT)   \cite{Brigham,Gold_Eng} of EV and current response is used  \cite{Review,DFT,DFT2}. After which, the investigated system parameters are estimated  by fitting the obtained IS with the  complex nonlinear least squares method (CNLS) \cite{MacdonaldR}.

Obviously, the nature of dynamic and non-reversible systems does not allow experiment replication for data accumulation with the following statistical noise-canceling (averaging signal technique). To study them the technological improvements such as shielding, increasing EV, varying geometry of electrodes or low-noise electronics usage are required for increasing signal to noise ratio (SNR). If the sample produces its own noise or when the mentioned above methods are impractical, the obtained raw IS is distorted by noises and interferences, which complexifies data analysis. FFT does not provide the noise-canceling option by itself and thus the results of CNLS can be unreliable. For our best knowledge, only  weighting method is currently used in this case for CNLS fit   \cite{Barsoukov,Lvovich,about_weight}. If this approach does not succeed, the IS data can not be interpreted and  no information about the studied system can be gained.

The progressive signal processing methods (Kalman filtering \cite{PRX_Kalman,PRApplied_Kalman}, adaptive filtering \cite{Stearns}, and other \cite{Gold_Eng}) allows extraction more information from high noise data.
The possibility of the adaptive filtering (AF) application for the IS processing has been previously noticed  \cite{Wang}. This approach can provide the solution of the problem mentioned above, due to the noise-canceling property of AF in the identification mode (learning)   \cite{Stearns}. 
Together with the usage of the broad frequency spectrum excitation signal, like in standard FFT method, decreasing the instrument influence on the sample and increasing time and frequency resolution   \cite{Denda_Eng,Denda_Deu,Review,DFT,DFT2} as high as it is principally possible. In addition the AF requires relatively small computation power and available memory, which is important for on-line applications and creating portable devices. The information on IS is stored in weight coefficients (WC), decreasing memory requirement for IS data storage and to increasing IS data transfer speed. Finally, after filter learning phase, AF opens the direct way for creating the digital model resembling the behavior of the investigated system, which is useful for its further simulation.

Despite the successful application of the AF algorithm for the IS treatment in Ref.  \cite{Wang}, no theoretical background has been given for this approach and no comparison with other methods has been done. Present study is devoted to elucidation of these aspects.

In this paper we develop the theory of AF application for EIS. We analyze the relationship between  the adaptive filter weight coefficients and the IS, derive the apparatus and weight functions of the AF method and theoretically justify its noise immunity. 
We experimentally show that the AF method is more accurate than the FFT, especially for high-noise raw data.

The paper is organized as follows. In subsection \ref{Model} we introduce the AF model employed and in subsection \ref{Basic relationsheeps} we provide derivation of the system immittance on the basis of AF parameters, namely weight coefficients. In section \ref{Experimental} we describe the experimental setup for measuring immittance.  In section \ref{Results} on the basis of the obtained experimental data we justify the noise immunity property of AF with respect to FFT. In section \ref{Bio-sensing} we demonstrate power of AF by its application for EIS-based in-vitro cell bio-sensing.
\section{Theory}
\subsection{Model}
\label{Model}
\begin{figure}[h]
\begin{center}
 \includegraphics[scale=0.45]{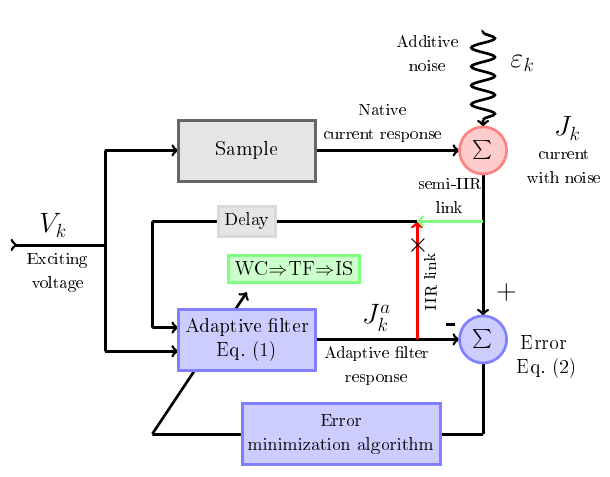} 
\end{center}
\caption{Processing the IS on the basis of AF method. Classical AF identification system with IIR corresponds for enabled IIR link (red) and disabled semi-IIR link (green), on the contrary this scheme describes semi-IIR model. If both links disabled, scheme describes FIR model.\label{fig:AF_Standard}}
\end{figure}

For processing the IS on the basis of AF method the investigated sample can be considered as a linear "black-box" (Fig. \ref{fig:AF_Standard}). 
The experimentally measured quantities are the exciting voltage $V_k$ being the input signal and current response $J_k$ being the output signal or in the terms of AF the desired signal. Index $k$ here is the time counter. The additive current noise $\varepsilon_k$ can also arise during measurements, due to the sample own noise, external noise or interfernces. We will now introduce $J_k^a$ as predictions of the adaptive filter:

\begin{equation}
  \label{Filter}
  J_k^{a}=\sum\limits_{j=1}^{\ell_d} d_j J_{k-j}^{a}+\sum\limits_{j=0}^{\ell_n} n_j V_{k-j},
\end{equation}
where $d_j$ and $n_j$ are weight coefficients (WC), and the maximum value between $\ell_d$ and $\ell_n$ is the filtering order. So, Eq. \eqref{Filter} describes the causal (without loss of generality) discrete  filter with infinite impulse response (IIR)  \cite{Gold_Eng}.
By adjusting the WC the following functional on the overall result of measurement is to be minimized
\begin{equation}
\label{Standard}
  \sum\limits_{k=\max(\ell_n,\ell_d)}^L \left| J_k^a-J_k\right|^2=\min,
\end{equation}
where $L+1$ is  the total number of collected samples.
The transfer function (TF) of Eq. \eqref{Filter} with the adjusted WC can be considered as an approximation of the sample admittance  \cite{Wang} as it will be shown below in \ref{Basic relationsheeps}.

It is well known, that the IIR filters are more flexible than the finite impulse response (FIR) filters. However the search of the WC values for IIR filters requires sufficient computation power and principal problem of the functional \eqref{Standard} local minimums exists \cite{Stearns}. For increasing flexibility on the one hand, and for simplifying WC search on another hand, in this paper we introduce the "semi-IIR" filter model. Namely, we replace in the right-hand part of Eq. (1) $J_k$ with $J_k^a$ before substituting it to Eq. (2). Thus the functional to be minimized takes a form
\begin{equation}
 \label{eq:Funtional}
\sum_{k=\max(\ell_n,\ell_d)}^L\left|J_k-\sum_{j=1}^{\ell_d} d_j J_{k-j}-\sum_{j=0}^{\ell_n} n_j V_{k-j}\right|^2=\min. 
\end{equation}

Eq. \eqref{eq:Funtional} describes operating principle of AF with $V_k$ for input signal and $J_k$ for the desired response and for AF prediction simultaneously. Note that beginning of summation from $j=1$ for $d_j$ in Eq. (3) prevents the trivial solution $d_0=1$, $d_j=0$ when $j>0$ and $n_j=0$.  The developed theory is also applicable for the FIR filters  \cite{Wang} characterized by fixing $d_j=0$ and $\ell_d=0$, for the non-causal FIR and the non-causal "semi"-IIR filter models.
\subsection{The relation between time- and frequency domain}

\label{Basic relationsheeps}

We shall now discuss the possibility of IS estimation with Eq. \eqref{eq:Funtional}. For simplifying our analysis, we neglect measurement noise $\varepsilon_k$. If the sample rate $f_0$ is high enough one can change the summation by $k$ in Eq. \eqref{eq:Funtional} with integration over data collecting interval $[-T/2, T/2]$.  On the basis of Parseval's identity one can write the following  
\begin{multline}
\label{Parsevals transition}
\int\limits^{T/2}_{-T/2} \left|J(t)-\sum_{j=1}^{\ell_d} d_j J(t-j\Delta t)-\sum_{j=0}^{\ell_n}  n_j V(t-j\Delta t)\right|^2 \df t=\\
=\int\limits^{+ \infty}_{-\infty} \left|J(t)-\sum_{j=1}^{\ell_d} d_j J(t-j\Delta t)-\sum_{j=0}^{\ell_n} n_j V(t-j\Delta t)\right|^2\Pi\df t=\\
=2 \pi \int\limits^{+ \infty}_{-\infty}\left|\left(1-\sum\limits_{j=1}^{\ell_d} d_j\varphi_j\right)J(f)-V(f)\sum_{j=0}^{\ell_n}n_j\varphi_j \right|^2*K  \df f,
\end{multline}
where  $\Pi$ is rectangle function in the interval $[-T/2,T/2]$, with Fourier image given by $\sinc$ function
\begin{equation}
K=\frac{\sin( \pi fT)}{\pi f},
\end{equation}
$\varphi_j(f)= \exp[2 \pi i\cdot  j(f/f_0)]$ is the Fourier image of the impulse delayed by $\Delta t=1/f_0$, $f$ is the frequency, asterisk $*$ denotes convolution, $V(f)$ and $J(f)$ are Fourier images of voltage and current, respectively.  Now we use Ohm's law $ J(f)=Y V(f)$, where $ Y $ is admittance of the system at frequency $f$, and omit $2 \pi$ multiplier. Thus from \eqref{Parsevals transition} one can write for the functional to be minimized:

\begin{multline}
\label{Admittance approximation}
\int\limits^{+ \infty}_{-\infty}\left[\left|\left( 1-\sum\limits_{j=1}^{\ell_d} d_j \varphi_j\right)Y-\sum_{j=0}^{\ell_n} n_j \varphi_j \right|^2 \right.\\
\left. \vphantom{\left( 1-\sum\limits_{j=1}^{\ell_d} d_j \varphi_j\right)}\times |V(f)|^2\right]*K \df f=\min.
\end{multline} 
Let us analyze Eq. \eqref{Admittance approximation}. It is simple to see that $K$ plays a role of the apparatus function, which defines the frequency resolution of the method $\Delta f$. The latter can be calculated as the distance between the first positive and the first negative roots of $K$, which can be estimated as 

\begin{equation}
\label{Resolution}
\Delta f = \frac{1}{T}.
\end{equation} 

If the data collection time $T$ is long enough, $K$ can be replaced with the $\delta$-function, which allows to omit the convolution in Eq. \eqref{Admittance approximation}. Henceforth we take convolution into account only as restriction on frequency resolution.

In these assumptions the squared magnitude of the EV image $|V(f)|^2$ is the weight function. Variation of its shape gives the possibility to control the influence of the selected frequency ranges  on the IS processing. 

According to Kotelnikov's theorem, the condition $\fb\leq f_0/2 $ must be met, where $\fb$ is the highest harmonic the exciting signal generator can produce. As a result, we obtain the following hierarchy of the characteristic setup frequencies:
\begin{equation}
\label{eq:hierarchy}
\Delta f \ll\fb \leq f_0/2.
\end{equation}
       
The most illustrative and practical is the case of exciting voltage with $|V(f)|=\const$ in the frequency band $f\leq \fb$. The excitation voltage types satisfying this condition are the white-noise, linear sweep and $\delta$-function signals. Thus Eq. (6) takes a form

\begin{equation}
\label{Admittance Levy approximation}
\int\limits^{+\fb}_{-\fb}\left|\left( 1-\sum\limits_{j=1}^{\ell_d} d_j \varphi_j\right)Y-\sum_{j=0}^{\ell_n} n_j \varphi_j \right|^2 \df f=\min.
\end{equation} 

The integrand for case of IIR filter usage can be obtained by dividing the integrand in Eq. \eqref{Admittance Levy approximation} by the squared absolute value of the expression in round brackets.

Eq. \eqref{Admittance Levy approximation} is in fact the Levy approximation \cite{Levy} for the admittnace $Y$ with $\varphi_j$ being the basis functions and one can write the following estimation of the admittance:

\begin{equation}
\label{Admittance}
Y \approx\left(\sum_{j=0}^{\ell_n} n_j \varphi_j \right) \left/ \left( 1-\sum\limits_{j=1}^{\ell_d} d_j \varphi_j\right) \right. . 
\end{equation}

One sees that the admittance $Y$ is directly derived on the basis of WC obtained by minimization of the functional in Eq. \eqref{eq:Funtional} with experimentally measured sequences of $V_k$ and $J_k$ substituted. AF method does not rely on FFT, which gives the advantage of memory economy. Since admittance estimation \eqref{Admittance} is derived from Eq. \eqref{Admittance Levy approximation} by heuristic manner,  there are no obvious preference to use admittance $Y$ or impedance $Z=1/Y$ for IS analysis.

In the special case of FIR filter (denominator in Eq. (9) is equal to 1) together with additional condition  $\fb=f_0/2$, WC $n_j$ are straightforwardly the Fourier coefficients of admittance. It means that any admittance, which can be expanded into Fourier series in the $\pm f_b$ interval (including ideal circuits and Warburg impedance) can be estimated by AF with any accuracy.
In practice the most actual case is $f_0\gg \fb $. Thus for low enough AF order ($\ell_d,\ell_n \ll f_0/\fb$) the $\varphi_j(f)$ arguments $2 \pi i\cdot  j(f/f_0)$ are much less than 1. It means that the exponential functions $\varphi_j(f)$ can be expanded into the Taylor series and Eq. \eqref{Admittance} becomes a rational function (standard Levy approximation) with coefficients being a linear combinations of $n_j$ and $d_j$. 

Typically the admittances of real systems are given by rational functions or expandable into Taylor and Fourier series. However, the straight algebraic relation between the parameters of the studied system (the ratings of resistors, capacitors and inductors the system is formed of) and AF WC is complicated. Therefore we suggest using Eq. \eqref{Admittance} for obtaining the frequency dependence of impedance. Due to the noise-cancelling property of AF, the yield of Eq. \eqref{Admittance} can be considered as a noise-free approximation of real IS. Then the system parameters can be can be derived by applying the standard methods to the obtained IS: algebraic method (AM) \cite{MacdonaldD}, geometric methods  \cite{Tsai} or CNLS \cite{MacdonaldR}.

The procedure of functional \eqref{eq:Funtional} minimization and noise-canceling property of AF are discussed in appendices \ref{Minimization} and \ref{About noise immunity}, respectively. 


\section{Experimental}
\label{Experimental}
The linear sweep-shape signal with 20 mV peak-to-peak amplitude, generated by  signal generator AKIP-3413-3 (AKIP, Russia, 2 channels), was used as excitation voltage. The frequency range was from 10 Hz to 40 kHz and sweep time was 500 ms. 

For $J_k$ measurements the ammeter (current-to-voltage converter) based on operational amplifier AD8606 (Analog Devices, USA) was assembled. It should be noticed that AD8606 has unity-gain frequency $f_s=10$ MHz and virtual inductance $L_V$ about several  mH can arise due to the used back-feed resistor $R_0=100$ k$\Omega$. The corresponding equation reads as $L_V = R_0/ (2 \pi f_s)$. 

\begin{figure}[h]
\includegraphics[scale=0.5]{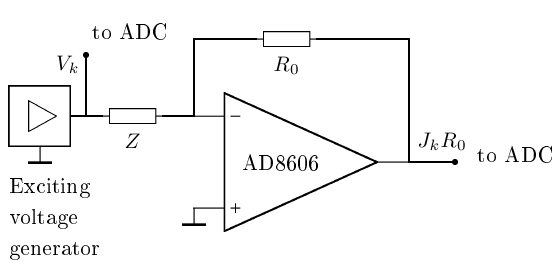}
\caption{Setup for IS measurement using the time-domain technique. Operational amplifier  is connected into current-to-voltage converter circuit. Here $Z$ is unknown sample impedance, $R_0$ is feed-back resistor. \label{Setup scheme}}
\end{figure}

The L-Card E20-10 ADC (L-Card, Russia, 10 MHz band, 4 channels, 12 bit) was used in this study to record the exciting voltage $V_k$ and current response $J_k$ from the ammeter output.  The data collection  time $T$ was 500 ms and sampling rate $f_0=500$ kHz, which yields 250 thousands for the collected samples number. One sees that the experimental setup parameters obey the required frequencies hierarchy Eq.\eqref{eq:hierarchy}  and constant EV spectrum assumption (see apendix \ref{About sweep}). The setup scheme is shown on Fig. \ref{Setup scheme}.

The serial RLC-circuit was used as a sample. The multimeter DT9205A (Resanta, Russia) yields the reference values of circuit elements $C_m=10$ nF, $R_m=250$ $\Omega$, the inductance factory rating is 15 mH. The measurements were conducted with various signal to noise rations (SNR). For decreasing SNR the interference source (transformer connected to the generating white noise-shaped signal second channel of AKIP) was placed near the RLC-circuit. This is a simulation of the usual experimental practice: any electronic device (microscope, computer, display, motors, transformers and power electronic) located in close proximity to the sample produces interference. This method resemble e.g. an electrical biological cell-substrate impedance sensing (ECIS) device  \cite{Geaver} or single cell EIS  \cite{Dittami} with microscope control (see section \ref{Bio-sensing}). SNR values were calculated as the relation between the noise-free $J_k$ std (EV output on, noise generator off) and the noise std (EV output  off, noise generator output on).

To obtain lower values of SNR the simulation with digital artificial noise was conducted. The values $\varepsilon_k$ were generated by Matlab  randn routine and added to current response $J_k$ obtained with the highest experimental SNR (30 dB). To achieve the statistics the data was collected over 4 experiments for each SNR for experimental  and artificial digital noise.

To implement the AF-based derivation of IS the introduced above semi-IIR model with $\ell_d$=49 and $\ell_n=101$ was used. To obtain the IS with FFT the Fourier images of $V_k$ and $J_k$ were calculated in Matlab. 

For further IS CNLS parameters fitting the Matlab code NELM based on Nelder-Mead simplex algorithm \cite{NM} was written. The latter may be obtained from the authors. The freeware LEVM program \cite{about_weight} by Ross Macdonald and Origin package were not suitable for this purpose due to impossibility to process large input data set and support CUDA technology for decreasing run-time. Initial values of $R$, $L$, and $C$ for CNLS fit were 256 $\Omega$, 19.4 mH and 9.207 nF, respectively.

\section{Results and discussion}
\label{Results}
\begin{figure*}
\centering
\Large{SNR=30 dB~~~~~~~~~~~~~~~~~~~~~~~~~~~~~~~~~~~~~~~~~~~~~~~SNR=3 dB}
\subfigure[]{\label{fig:Noise_immunity_R_30}\includegraphics[scale=0.3]{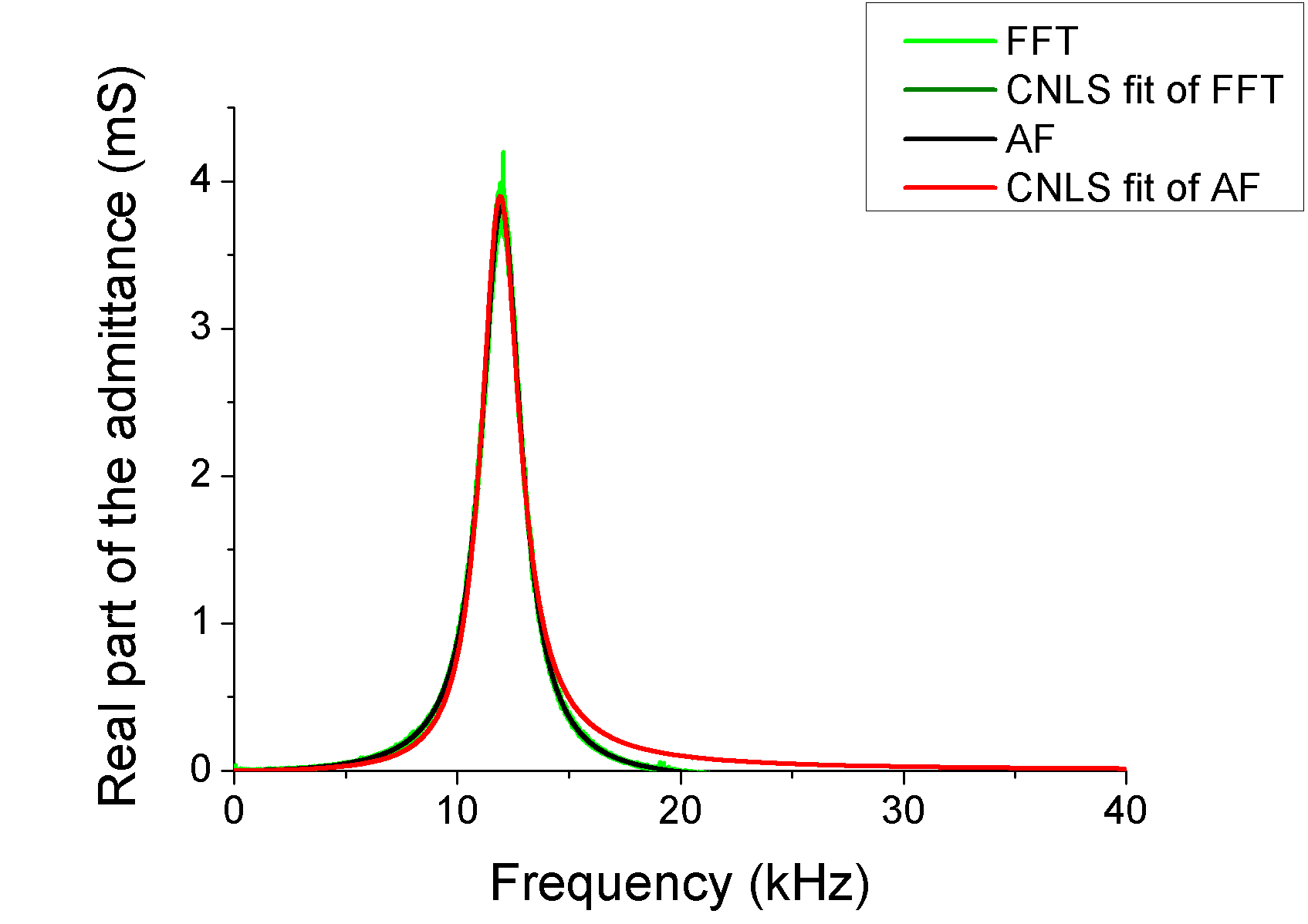}}\subfigure[]{\label{fig:Noise_immunity_R_3}\includegraphics[scale=0.3]{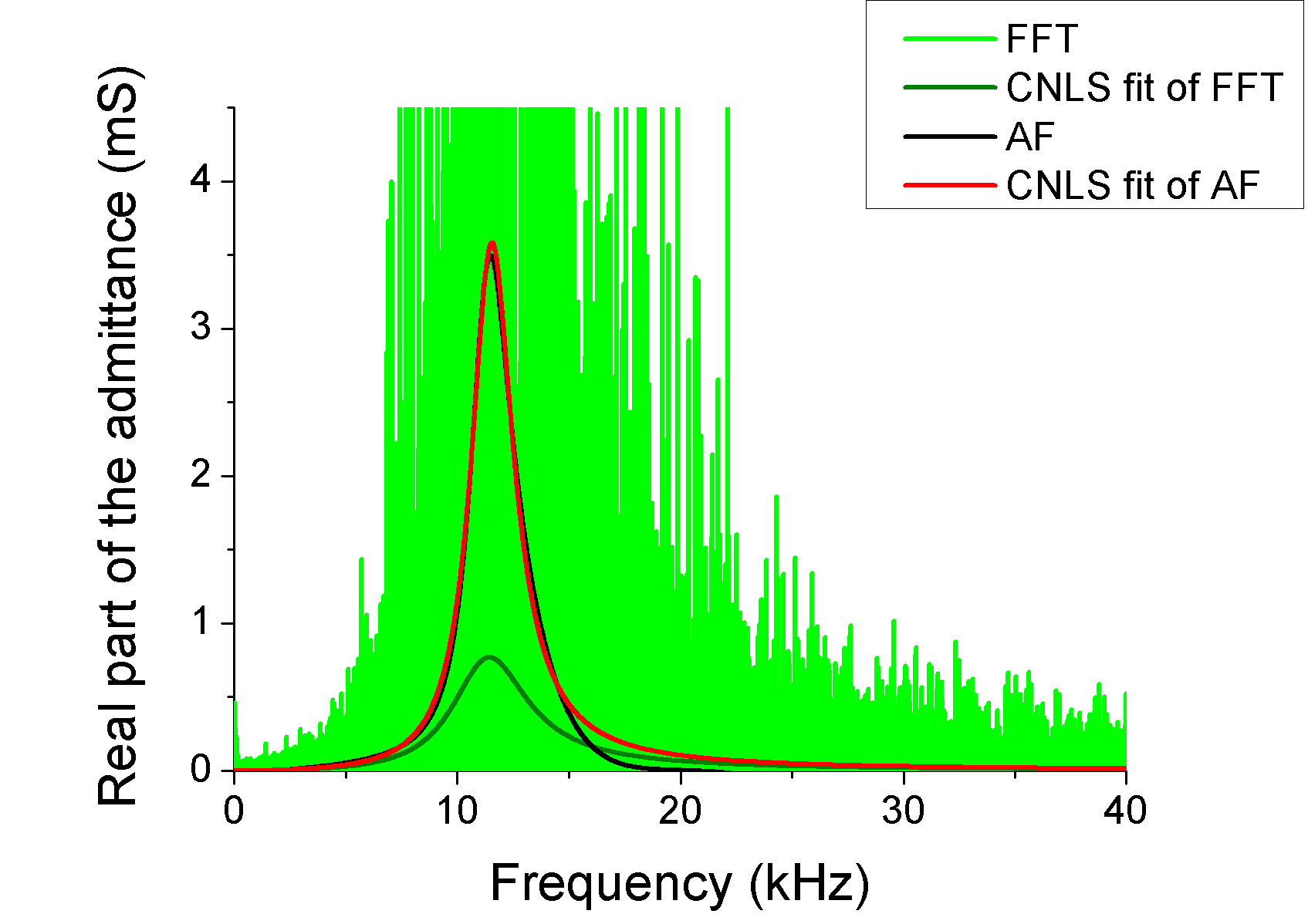}}
\subfigure[]{\label{fig:Noise_immunity_I_30}\includegraphics[scale=0.3]{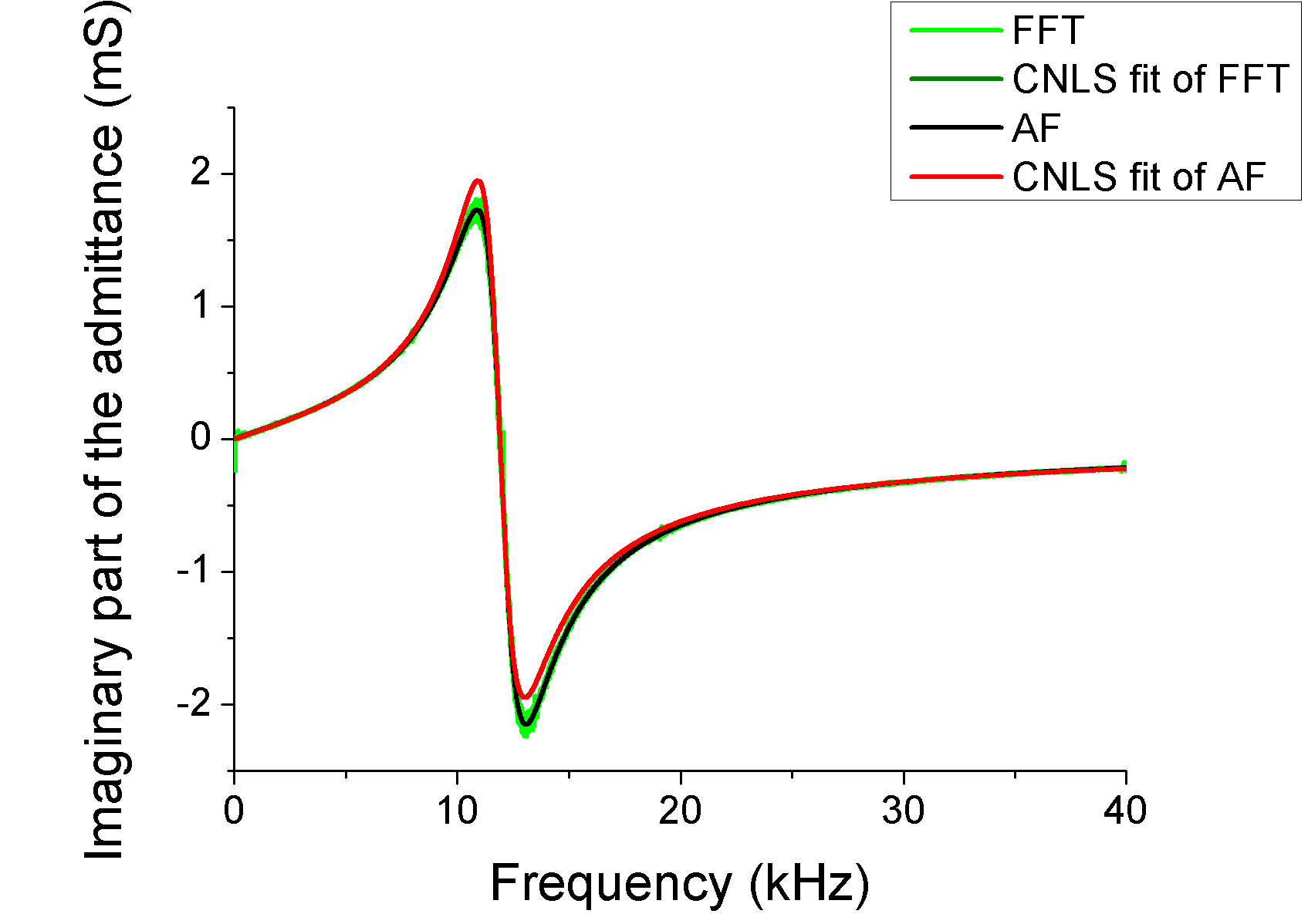}}\subfigure[]{\label{fig:Noise_immunity_I_3}\includegraphics[scale=0.3]{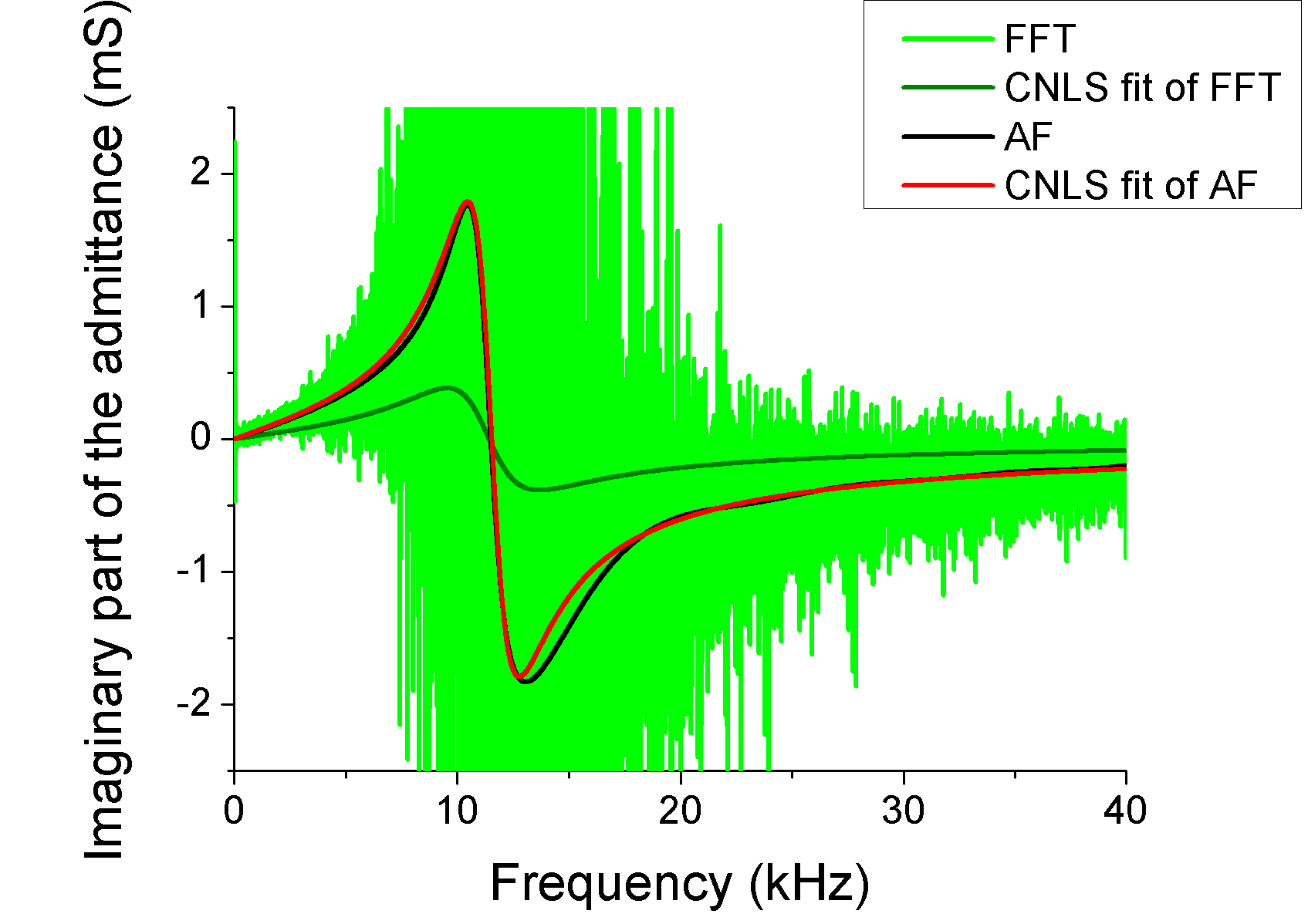}}
\subfigure[]{\label{fig:Noise_immunity_Lotus_30}\includegraphics[scale=0.3]{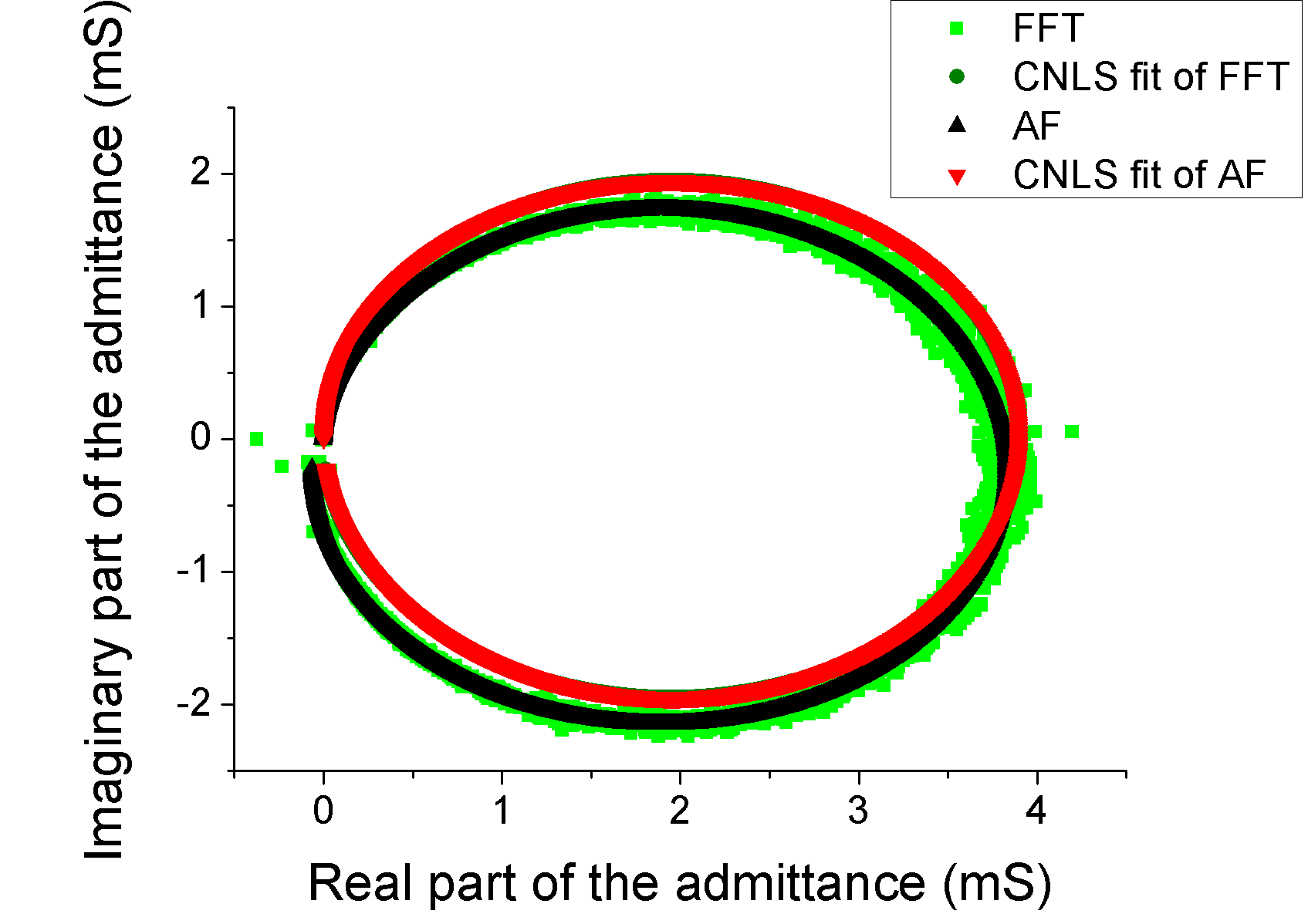}}\subfigure[]{\label{fig:Noise_immunity_Lotus_3}\includegraphics[scale=0.3]{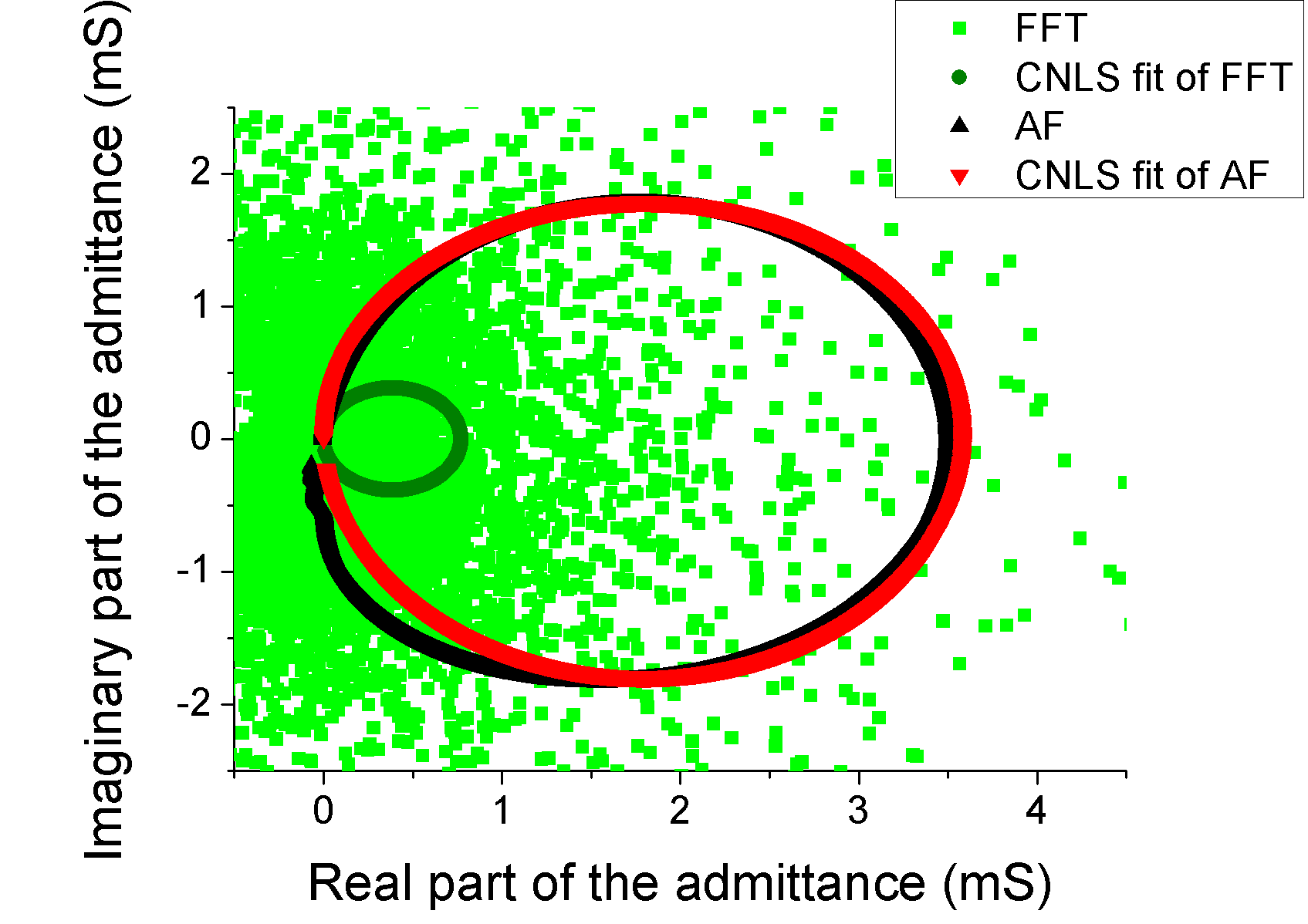}}

\caption{Frequency dependencies of admittance obtained with AF and FFT for the high and the low experimentally studied SNR. Panels (a), (c), and (e) correspond to SNR = 30 dB and show the admittance real and imaginary part spectra and the Nyquist plot, respectively. Panels (b), (d), (f) correspond to SNR = 3 dB. Light green color is used to show the raw admittance spectrum obtained by FFT and dark green color is used to show the CNLS fit of the corresponding spectra. The admittance spectra obtained by AF is shown with black color and its CNLS fit is shown by red color. The CNLS fits for AF and FFT on panels (a), (c), and (e) are overlapping.
 \label{fig:Noise_immunity}}
\end{figure*}

\begin{figure*}
\centering
\Large{Experiment~~~~~~~~~~~~~~~~~~~~~~~~~~~~~~~~~~~~ Simulation}
\subfigure[]{\label{fig:Noise_immunity_R_E}\includegraphics[scale=0.3]{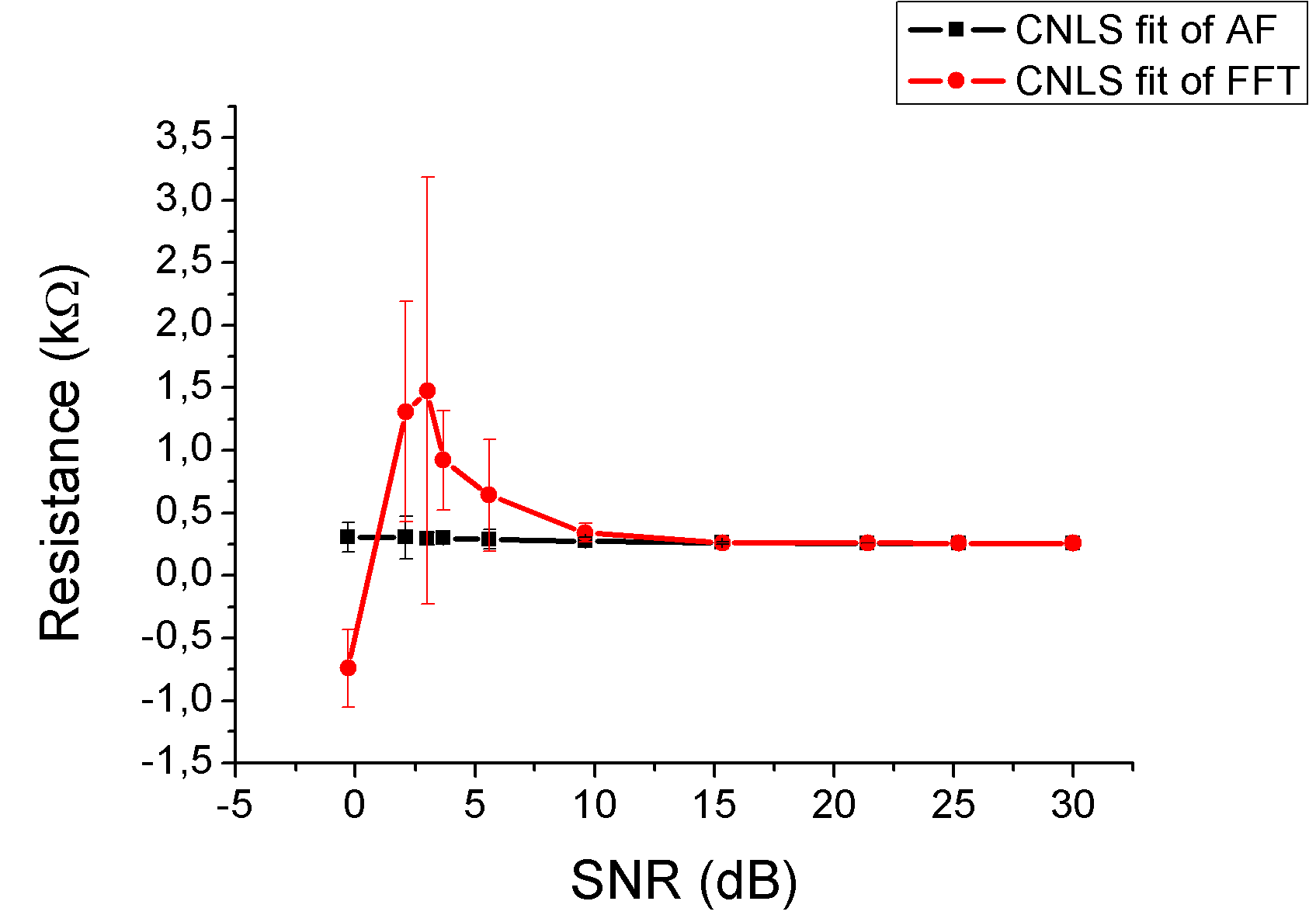}}\subfigure[]{\label{fig:Noise_immunity_R_S}\includegraphics[scale=0.3]{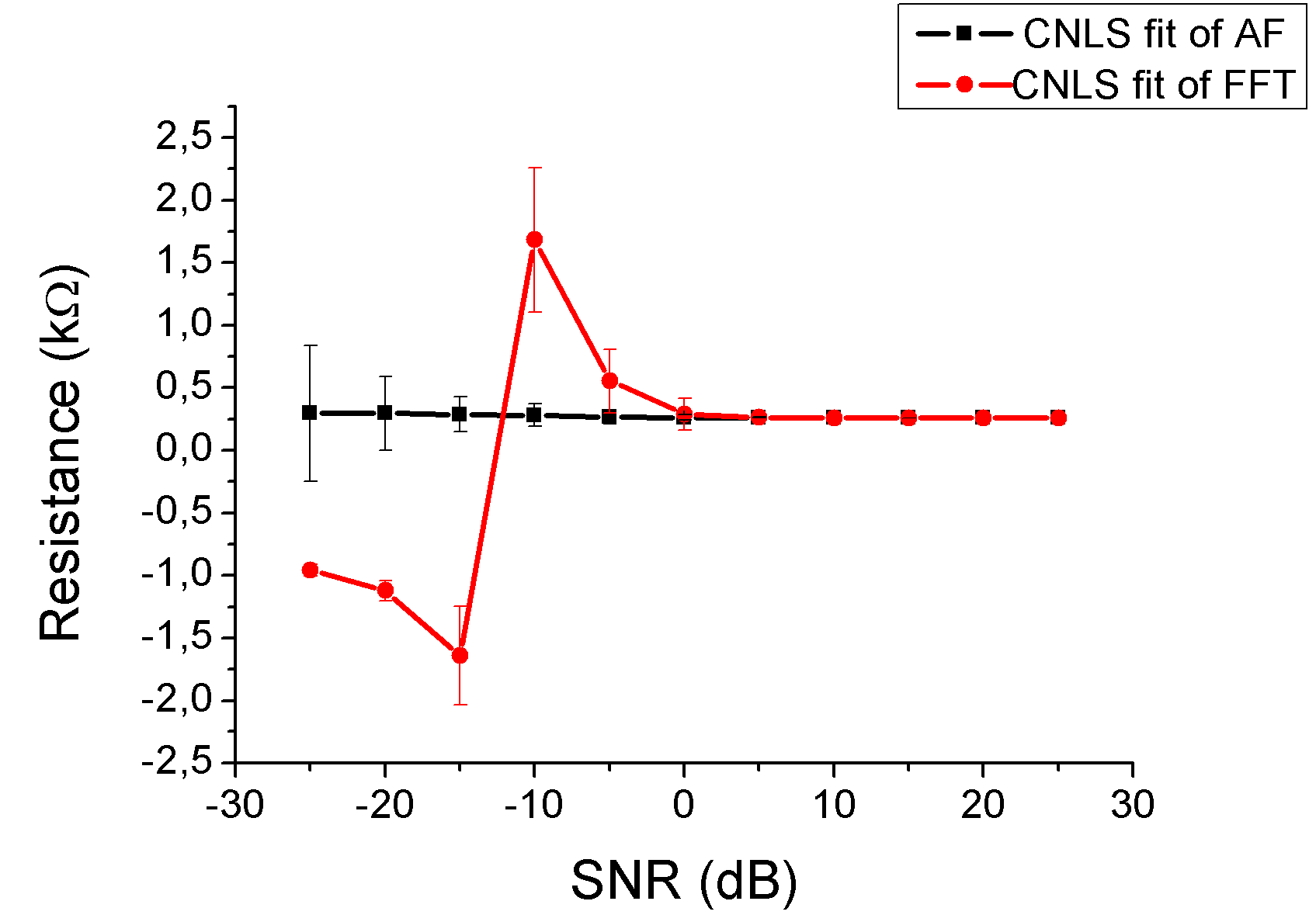}}
\subfigure[]{\label{fig:Noise_immunity_C_E}\includegraphics[scale=0.3]{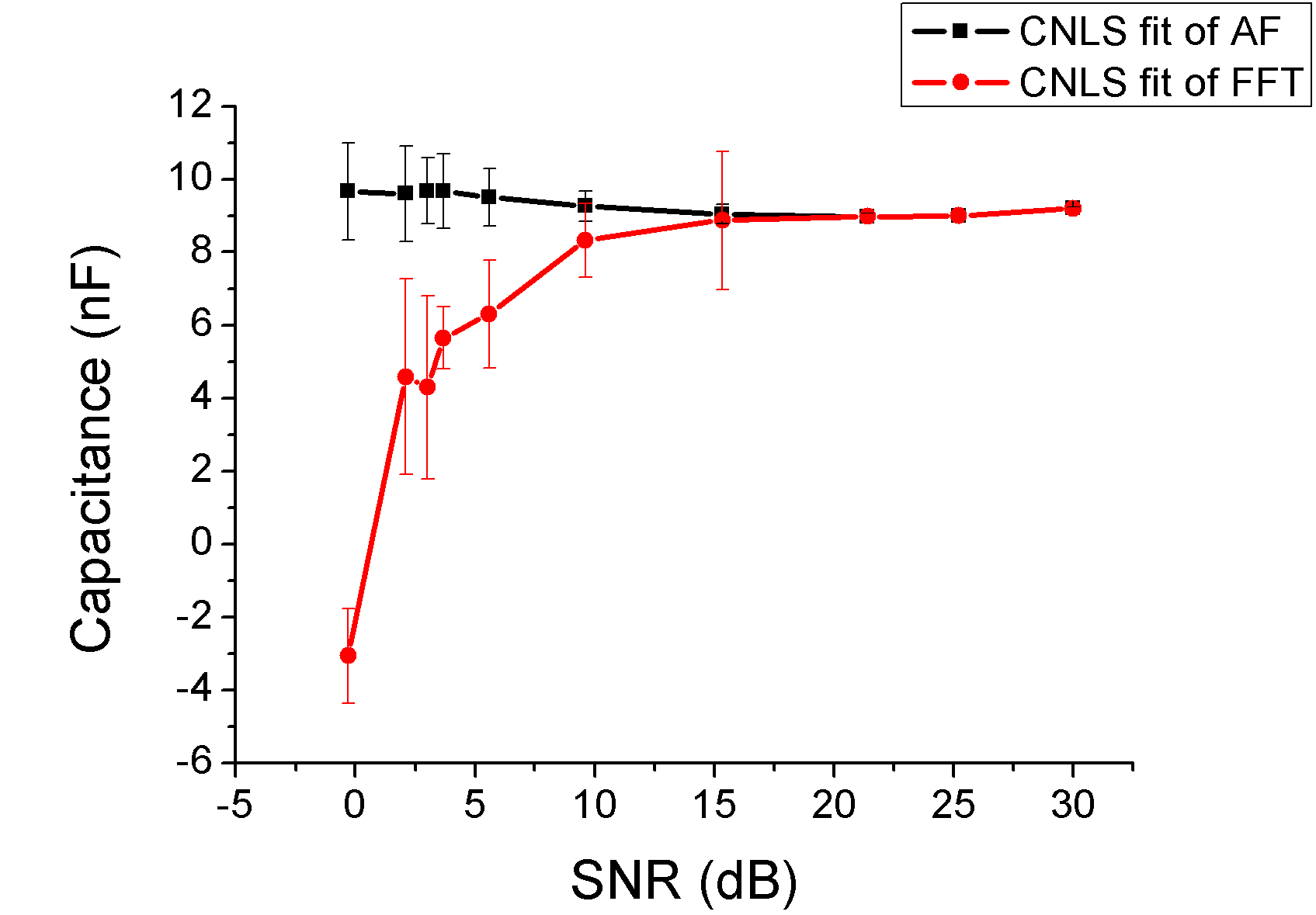}}\subfigure[]{\label{fig:Noise_immunity_C_S}\includegraphics[scale=0.3]{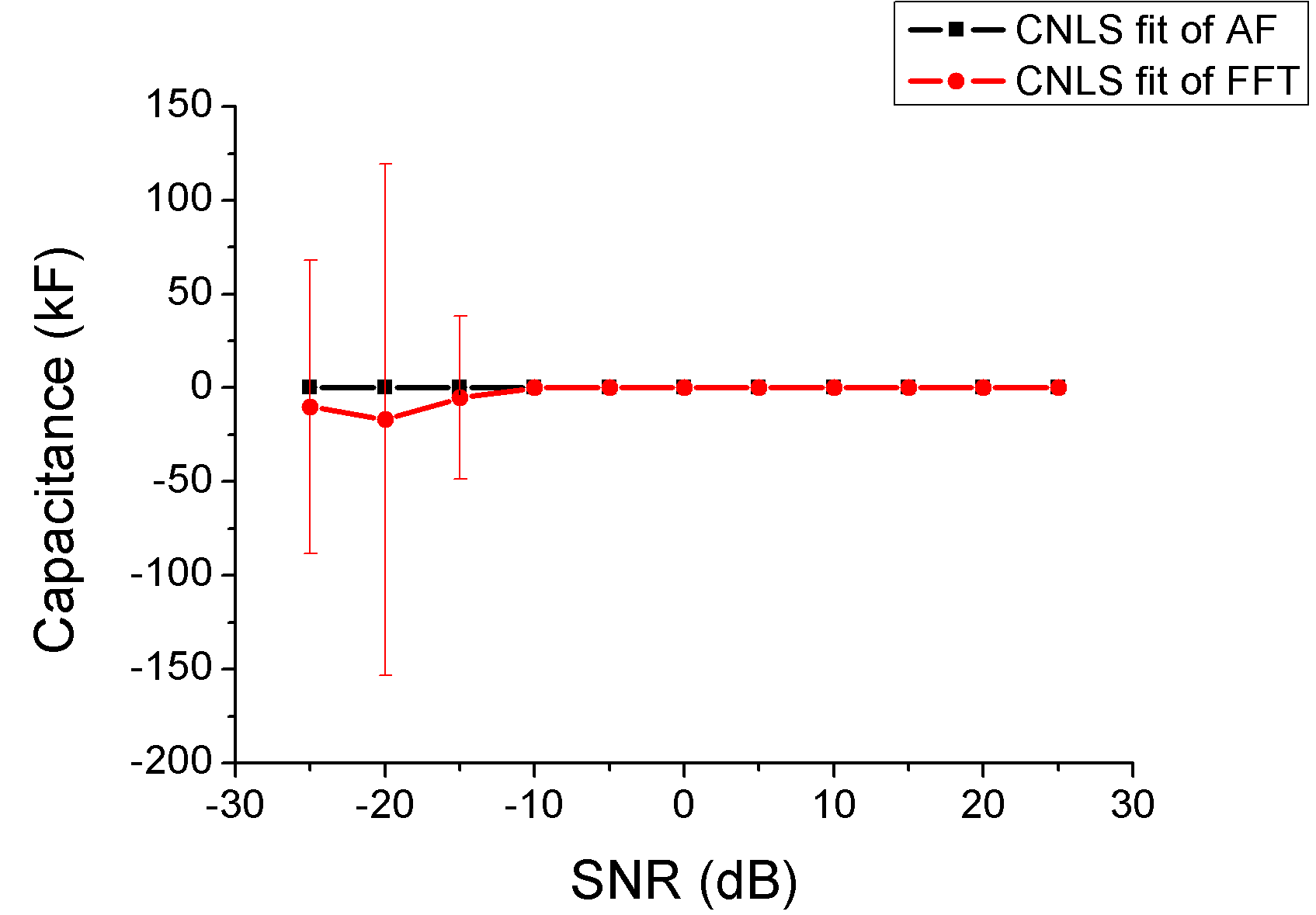}}
\subfigure[]{\label{fig:Noise_immunity_L_E}\includegraphics[scale=0.3]{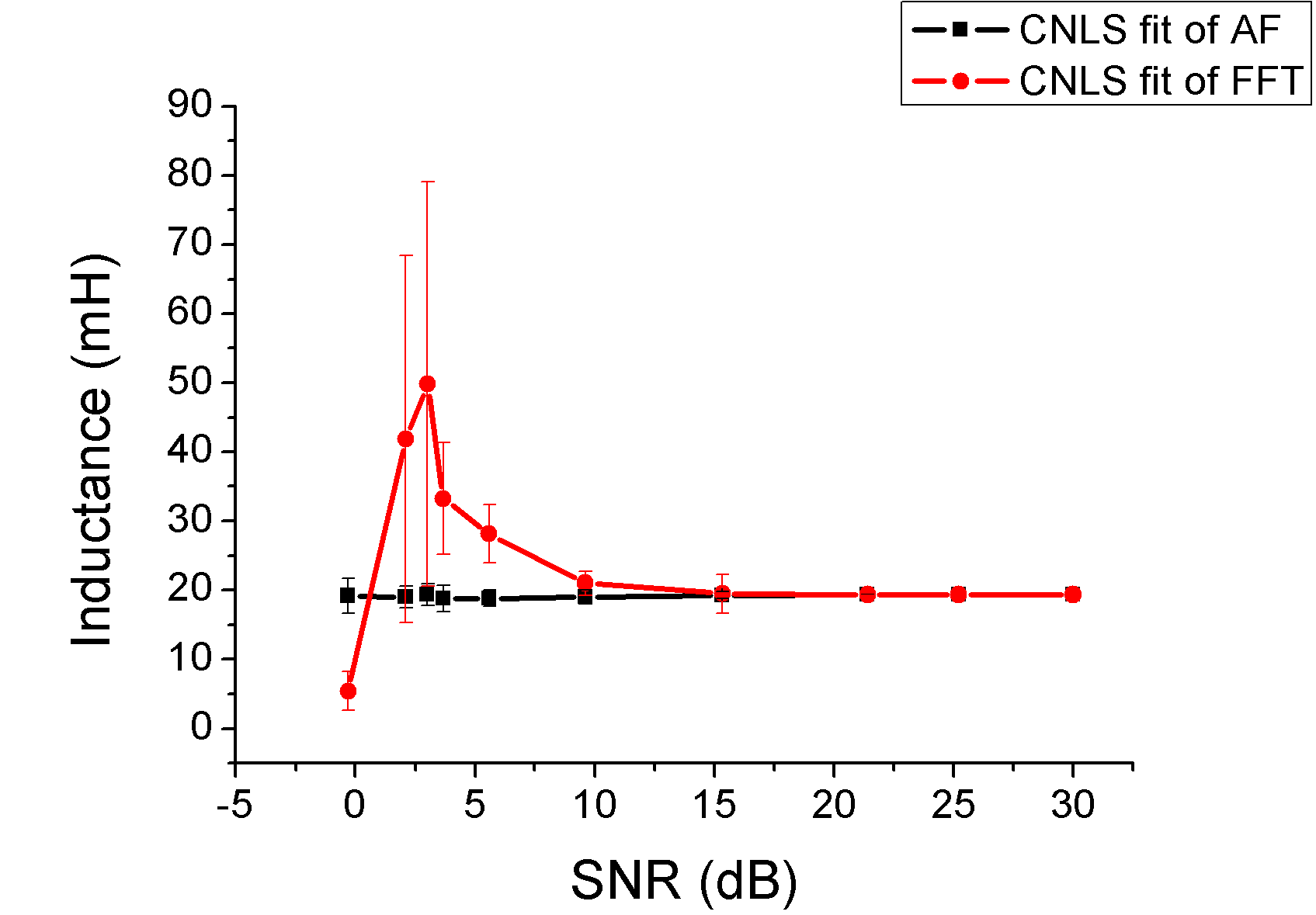}}\subfigure[]{\label{fig:Noise_immunity_L_S}\includegraphics[scale=0.3]{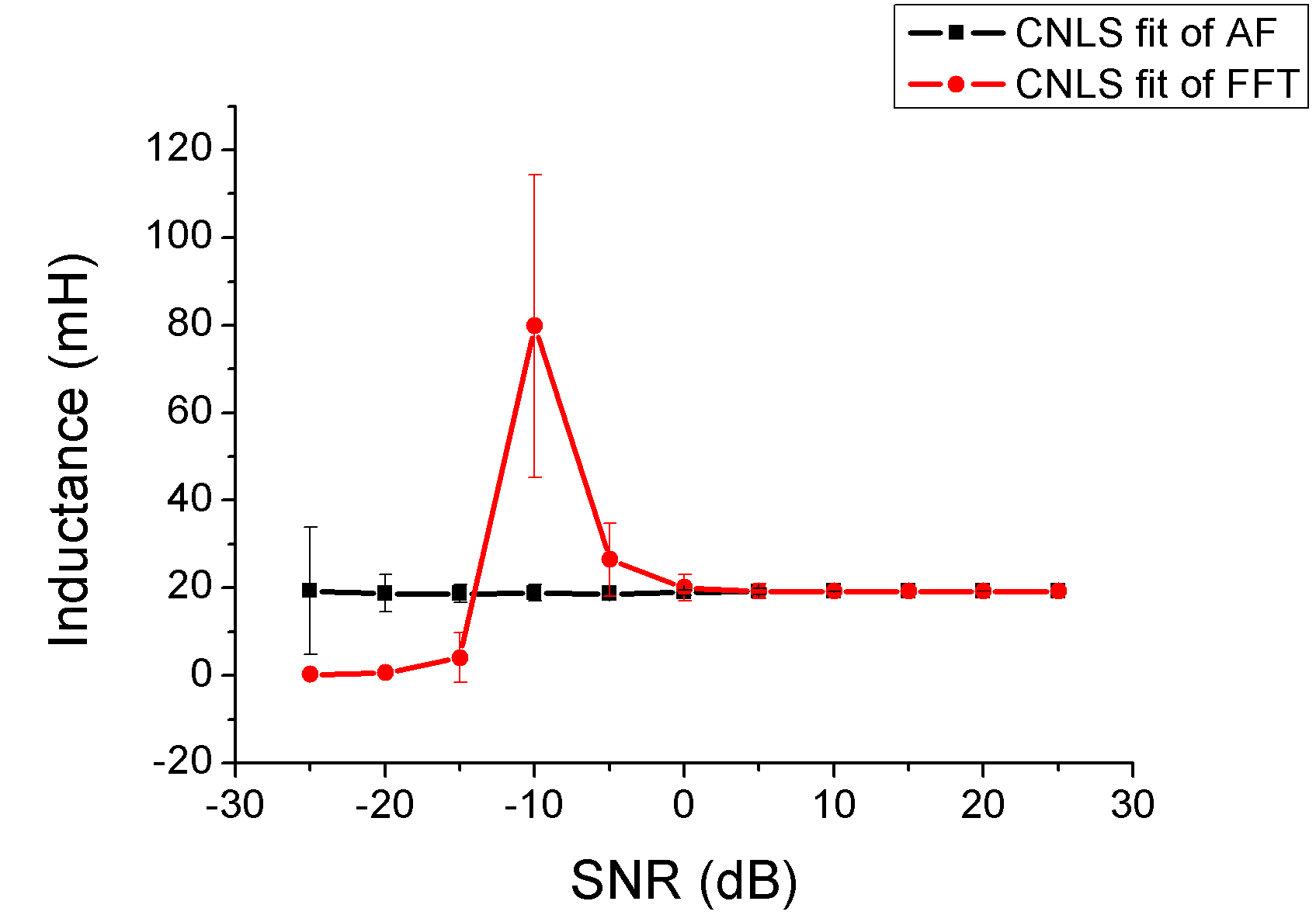}}
\caption{\label{fig:Noise_immunity_Experiment_and_Simulation} Dependence of the obtained circuit elements ratings on signal to noise ratio exhibiting the AF advances in noise immunity with respect to FFT. Black squares are for the results of CNLS fit of IS obtained by FFT. Red circles denote the circuit elements ratings obtained by AF. The statistics over 4 experiments for each SNR allows adding the error bars (99.9\% reliability).
Panels (a) and (c), (e) are for experimentally generated noise for resistance, capacitance and inductance, respectively. Panels (b), (d) and (f) are for digital artificial noise for resistance, capacitance and inductance, respectively. AF gives correct results for all studied SNR for both experimental noise and noise modeling. FFT fails at 5 dB for experimental data, and at 0 dB for noise modeling. }
\end{figure*}

  The dependence of circuit element ratings CNLS fit on SNR for IS obtained by AF and FFT methods are presented on Fig. \ref{fig:Noise_immunity_Experiment_and_Simulation}. In Fig. \ref{fig:Noise_immunity} the spectra and the corresponding CNLS fits for the 30 dB and the 3 dB SNR are shown. Corresponding CNLS fit results are listed in Table \ref{tab:Comparsion}. One sees that AF-method is robust with respect to decreasing SNR.

\begin{table*}
\caption{\label{tab:Comparsion} RLC-circuit parameters estimation with 99.9\% confidence intervals  for different methods and SNRs. Last columns list the STD between CNLS fit of FFT yield and obtained with AF spectrum with respect to FFT.}
\begin{ruledtabular}
\begin{tabular}{ccccccc}
 SNR, dB &  Method     &  $R$, $ \Omega$  &  $C$, nF &   $L$, mH& Residual type& Residual  RMS, $\mu$S\\ \hline
30 & AF $\&$ CNLS &$256.7 \pm0.4$ &$9.209 \pm0.007$ &$19.36 \pm0.02$&AF vs. FFT& 30\\
   &FFT $\&$ CNLS&$256.7 \pm0.3$ &$9.207 \pm0.006$ &$19.360 \pm0.007$&CNLS vs. FFT& 110\\ \hline
 3 & AF $\&$ CNLS &$290 \pm50$ &$9.7 \pm0.9$ &$19 \pm2$&AF vs. FFT& 3000\\ 
   & FFT $\&$ CNLS &$1000 \pm 2000$ &$4 \pm3$ &$50 \pm30$&CNLS vs. FFT& 2000\\ \hline \hline
30 & AF $\&$ AM&$262.1 \pm0.4$ &$9.37 \pm0.02$ &$19.04 \pm0.04$& --\\ 
 3 &           &$300 \pm60$ &$9 \pm1$ &$21 \pm3$& --\\
\end{tabular}
\end{ruledtabular}
\end{table*}

At low SNR the spectra obtained by FFT are drastically distorted, which leads to incorrect estimation of $R$, $L$ and $C$ in CNLS fit. The FFT \& CNLS fit fails at low SNR even in the case, where the correct ratings obtained at high SNR are used as initial fitting parameters. The discrepancy in capacitance value is especially high. No significant difference between unit, proportional, modulus and $|V(f)|^2$ weighting (see Table I in ref.  \cite{about_weight}) on the FFT \& CNLS fit results was observed. Contrary to FFT, the IS obtained by AF method do not degrade with SNR decreasing, and $R$, $L$ and $C$ ratings calculated from them stay intact. We relate this effect to the noise-canceling property of AF. Eliminating the uncorrelated with exciting voltage noise from IS prevents its influence on the circuit parameters  estimation. Being the simplest artificial intelligence, AF automatically recognizes useful signal  in the time domain data and separates it from the signal distorted by background noise.

It is not surprising that the confidence intervals (CI) of circuit elements ratings for both AF and FFT methods increase with decreasing SNR. For AF CI grow with decreasing SNR slower than in the case of FFT. Lower spectrum AF vs. FFT STD in the high SNR case is also not surprising because as can be seen from FFT data in Fig. \ref{fig:Noise_immunity_Lotus_30} RLC-circuit IS is slightly distorted at high frequencies due to operational amplifier nonideality. AF semi-IIR model is more flexible than RLC-circuit model so the first one keeps this effect. 

Interestingly, in the case of low SNR the residual error between IS obtained by FFT and the spectra obtained with AF is higher than that of FFT spectra and their CNLS fit. So, the least mean squares criterion is not adequate for low SNR and highly distorted IS. To the best of our knowledge, there is no proof of the CNLS noise immunity and present results ascertain this figure. 

Smooth nature of the IS obtained by AF gives the possibility to use simple algebraic methods for circuit element ratings estimation, which gives significant advantage in computation power and eliminate human's factor. For RLC circuit the elements ratings can be expressed via impedance magnitude minimum position frequency $f_{\min}$, impedance magnitude minimum value $|Z(f_{\min})|$, and frequency of the admittance imaginary part maximum  position $f_{\max}$ as $R=|Z(f_{\min})|$, $L=Rf_{\max}/[2\pi(f_{\max}^2-f_{\min}^2)]$ and $C=1/[L(2 \pi f_{\min})^2]$.  The results of AM fit are given in the last two rows of Table \ref{tab:Comparsion} for comparison with the results of CNLS. The element rating CI  obtained by AM are not overlapped with CI for high SNR CNLS fits, because the AM use only few points from the IS at middle frequencies, and mentioned  high-frequency systematic distortion (ADC nonideality) does not affect them.
\section{Practical example. Application for bio-sensors}
\label{Bio-sensing}
In this section we present a practical example of bio-sensing, in which AF gives significant advantage with respect to FFT. 
\begin{figure}[h!] 
\centering
\subfigure[]{\label{fig:Control}\includegraphics[scale=1.5]{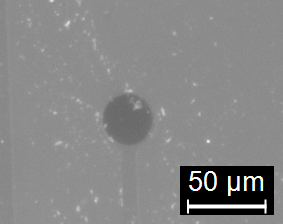}}
\subfigure[]{\label{fig:Cell}\includegraphics[scale=1.5]{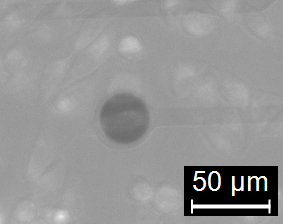}}
\subfigure[]{\label{fig:Cell_loki}\includegraphics[scale=0.3]{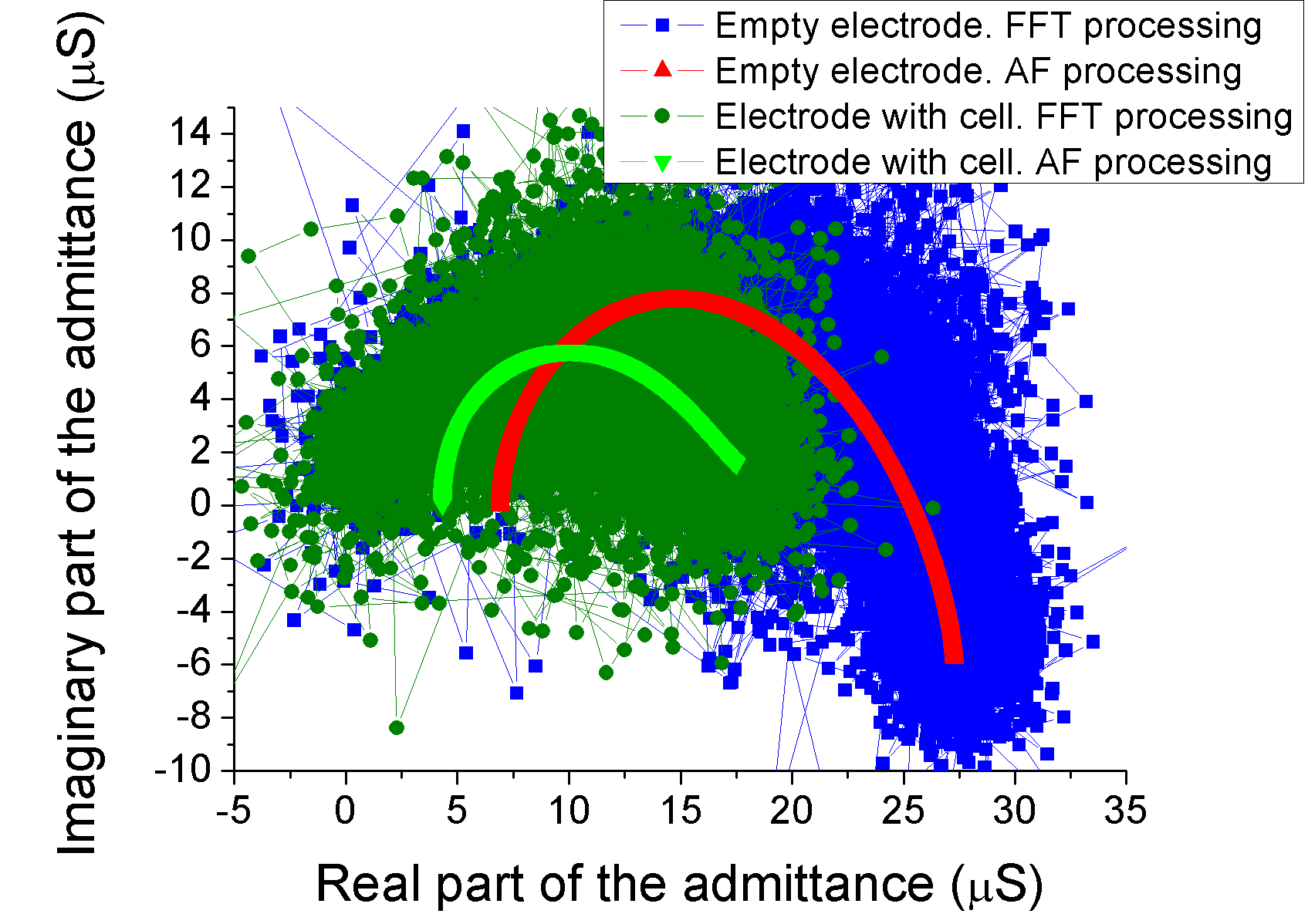}}
\caption{\label{fig:Cells} Application AF for bio-sensing EIS. (a) Empty electrode; (b) Electrode covered by cell. Diameter of electrodes is 30 $\mu$m; (c) Nyquist plot of obtained IS. Red and light green color denote the obtained by AF IS of the empty and the covered by cell electrodes, respectively. Blue and dark green color denote the obtained by FFT IS of the empty and the covered by cell electrodes, respectively. One sees that AF IS is more informative with respect to FFT IS, for example, it can be seen and agreement with Giaever-Keese model \cite{Geaver}.} 
\end{figure}
The investigated system was multielectrode array MEA 200/30 (Multi Channel Systems GmbH, Germany, 30 $\mu$m electrode diameter)  covered by HeLa cells in phosphate buffered saline (Biolot, Russia) in vitro  under microscope study. HeLa cells were obtained from the Bank of cell cultures of the Institute of Cytology of the Russian Academy of Sciences. IS was measured between large rectangle  reference electrode (50 $\mu$m$\times 250\, \mu$m, not presented on photographs) and empty electrode (Fig. \ref{fig:Control}), and between reference electrode and electrode covered by a single cell (see Fig. \ref{fig:Cell}). For IS measurement was used described above setup. Feedback resistor rating was 1 M$\Omega$, all other parameters was same as for RLC-circuit study. Results are presented on Fig. \ref{fig:Cell_loki}.

One can see, that IS obtained by FFT is distorted by interference from microscope, and no useful information can be achieved from it. Contrary, IS obtained by AF method is noise-free and more informative, for example, bright green and red arcs origins (their left ends) are displaced as predicted by Giaever-Keese model \cite{Geaver}. 

\section{Conclusion and Outlook}
In this study we have developed the theory of AF IS processing and compare this approach with FFT. The main outcome is that processing of time domain impedance data collected at low SNR conditions with FFT yields distorted and inadequate IS, but the IS obtained with AF method are robust with respect to noises even at negative SNR. The developed AF-based software for IS data processing can be obtained from the authors.

The developed AF-based approach makes impedance/admittance spectroscopy much more sensitive and brings it to the new level in all areas: material science, biophysics, electronic devices characterization and quality control in semiconductor industry. Moreover, the time domain impedance data ($V_k$ and $J_k$ sequences), which can not be interpreted with FFT approach, can be analyzed with AF-based one  with higher probability of success.

This technique can find practical use in studying dynamic and nonreversible systems under the low SNR conditions. It is especially important for biological systems, because their investigation requires usage of the low excitation voltage and current, which results in decreasing of signal-noise ratio. Portable biosensors based on the developed method will be robust to the external interferences and noises. Moderate  computation power and memory requirements also make AF appropriate for employing in this area. 

\begin{acknowledgments}
To Y. Well and H. Child. 

We would like to thank A.V. Nalitov, O.I. Utesov, I. N. Terterov and A. L. Chernev for useful criticism and N.A. Knyazev for his help in biological experiment describing.
\end{acknowledgments}

\appendix

\section{Functional minimization}
 \label{Minimization}

On can see that Eq. \eqref{Filter} can be written in the terms of the Toeplitz matrices

\begin{widetext}
\begin{equation}
\label{wide equation}
L-\max(\ell_n,\ell_d)\left\lbrace\vphantom{\begin{matrix}
 V_{\ell_n}&...& V_{2}& V_{1} &V_{0}  \\
 V_{\ell_n+1}&...& V_{3}& V_{2}&V_{1} \\
 \hdotsfor{5}\\
 V_{\ell_n+r}&...& V_{r+2}& V_{r+1}&V_{r} \\
  \hdotsfor{5}\\
  \hdotsfor{5}\\
    \hdotsfor{5}\\ 
    V_{L}& ... & V_{L-\ell_n+1} & V_{L-\ell_n}&V_{L-\ell_n} 
\end{matrix}}\right.\left|\left|\overbrace{\begin{matrix}
 V_{\ell_n}&...& V_{2}& V_{1} &V_{0}  \\
 V_{\ell_n+1}&...& V_{3}& V_{2}&V_{1} \\
 \hdotsfor{5}\\
 V_{\ell_n+r}&...& V_{r+2}& V_{r+1}&V_{r} \\
  \hdotsfor{5}\\
  \hdotsfor{5}\\
    \hdotsfor{5}\\ 
      \hdotsfor{5}
\end{matrix}}^{\VH}\right|\right.
\left.\left.\overbrace{\begin{matrix}
 J_{\ell_d}& ... & J_2 & J_1&J_0  \\
 J_{\ell_d+1}& ... & J_3 & J_2&J_1  \\
 \hdotsfor{5}\\
J_{\ell_d+r}& ... & J_{r+2} & J_{r+1}&J_{r} \\
  \hdotsfor{5}\\
  \hdotsfor{5}\\
    \hdotsfor{5}\\ 
      \hdotsfor{5}
\end{matrix}}^{\JH}\right|\right|
\cdot \overbrace{\begin{Vmatrix}
 n_0 \\
 n_1\\
 n_2\\
...\\
 n_{\ell_n}\\
 d_1 \\
 d_2\\
 d_3\\
...\\
d_{\ell_d}
\end{Vmatrix}}^{\vec{w}}=
\overbrace{\begin{Vmatrix}
 J_{\ell_d+1} \\
 J_{\ell_d+2} \\
 J_{\ell_d+3} \\
 ... \\
 J_{\ell_d+r}\\
 ...\\
...
\end{Vmatrix}}^{\vec{J_d}},
\end{equation}
\end{widetext}
where $\VH$ and $\JH$ is Toeplitz matrices  generated by $V_k$ and  $J_k$ sequences respectively, $\vec{w} $ is a vector obtained by $n_j$ and $d_j$  WC concatenation,  and $\vec{J_d}$ is desired current response. In matrix notation the problem Eq. \eqref{eq:Funtional} takes a form 

\begin{equation}
\label{Problem is Toeplitz matrices terms}
\left(\begin{Vmatrix}
\VH \JH
\end{Vmatrix}
\vec{w}-\vec{J_d}\right)^2=\min
\end{equation}

It is well known, that solution of the Eq. \eqref{Problem is Toeplitz matrices terms} can be reduced to solving the linear system
\begin{equation}
\begin{Vmatrix}
\label{Normal equations}
\VH^T\\ \JH^T
\end{Vmatrix}\begin{Vmatrix}
\VH \JH
\end{Vmatrix}\vec{w}=
\begin{Vmatrix}
\VH^T\\ \JH^T
\end{Vmatrix}\vec{J_d}.
\end{equation}

Thus the functional Eq.\eqref{eq:Funtional} minimization problem leads to the $(\ell_n+\ell_d+1)\times (\ell_n+\ell_d+1)$ linear system on vector $\vec{w}$ of WCs from which the admittance can be directly obtained with Eq. \eqref{Admittance}. To prevent any misunderstanding related with the existence of various physical dimensions in Eq. \eqref{Normal equations} we suggest the following normalization. The impedance $Z$ should be normalized to feedback resistor $R_0$ and all voltage inputs for ADC should be normalized to ADC range (see Fig. \ref{Setup scheme}). This procedure makes all elements of $\VH$, $\JH$, $\vec{w}$, and $\vec{J_d}$ dimensionless.

\section{Noise immunity of AF}
\label{About noise immunity}
Now we will use normalization notation mentioned above and  explain the noise-cancelling property of AF in the identification mode. To do so, we introduce the Toeplitz matrix $\EH$ generated by current noise $\varepsilon_k$ sequence in the same manner as $\JH$ is generated by $J_k$ and add $\EH$ to $\JH$. Thus Eq. (3) transforms to
\begin{equation}
\label{System in the Toeplitz matrices terms}
\left(\begin{Vmatrix}
\VH^T \VH && \VH ^T \JH\\
\JH ^T \VH && \JH ^T \JH\\
\end{Vmatrix}+E_L\right)\vec{w}^*=\begin{Vmatrix}
\VH^T\vec{J_d}\\
\JH^T\vec{J_d}
\end{Vmatrix}+E_R,
\end{equation}
where 
\begin{equation}
E_L=\begin{Vmatrix}
0 && \VH^T\EH\\
\EH^T\VH && \EH^T \JH+ \JH^T\EH+\EH^T\EH \\
\end{Vmatrix},
\end{equation}
\begin{equation}
E_R=\begin{Vmatrix}
\VH^T\vec{\varepsilon}\\
\EH^T \vec{J_d}+ \JH^T\vec{\varepsilon}+\EH^T\vec{\varepsilon} \\
\end{Vmatrix}
\end{equation} 
are perturbations, $\vec{w}^*$ is perturbed WC vector. If noise  $\varepsilon$ is uncorrelated with  EV and current then  all products  in  $E_R$ and $E_L$ with  $\VH$ and $\JH$ are equal to zero and perturbation depends only on the noise correlation function. Moreover, if noise is auto-correlated, then $E_R=0$ and $E_L$ is a diagonal matrix with $\ell_d$ non-zero elements. $L_2$-norm of $E_L$ in this case is noise mean square value $\langle \varepsilon^2 \rangle$.
In the basis of standard approach for linear system error estimation \cite{Verzbitskii_Eng,Rice_eng,Golub} one can write
\begin{equation}
\label{residual}
 \frac{|\vec{w} -\vec{w}^*|}{|\vec{w}|} \leq \frac{\langle \varepsilon^2 \rangle \nu}{A},
\end{equation}
where $\nu$ and $A$ are condition number and  $L_2$-norm of main matrix in Eq.\eqref{Normal equations}, respectively. The fact that $\langle \varepsilon^2 \rangle$ is independent on the filtering order allows one to obtain the following estimation for root mean square of the relative error (RMSE) in  WC from Eq.\eqref{residual}:
\begin{multline}
\label{WC error}
\mbox{RMSE}= \frac{1}{|\vec{w}|}\sqrt{\frac{\sum\limits_{j=0}^{\ell_n+\ell_d} (\vec{w}_j -\vec{w}^*_j)^2}{\ell_n+\ell_d+1}} \leq \frac{\langle \varepsilon^2 \rangle \nu}{A\sqrt{\ell_n+\ell_d+1}}.\\ 
\end{multline}
In the most favorable case $\nu=1$ and RMSE of WC falls off as $1/\sqrt{\ell_n+\ell_d+1}$, which is a reminiscent of classical statistics law for signal averaging technique. If  FIR model is used there are  no influence of noise on WC (compare with ex. 6 on p. 226 in ref. \cite{Stearns}). 

It should be noticed, that linear dependency of columns in $ \begin{Vmatrix} \VH \JH \end{Vmatrix}$ leads to increasing of $\nu$ and, as corollary, to  instability of the system  \eqref{Normal equations}.
Such linear dependency can arise in the case of short periodic EV or in the case of pure resistance-type impedance. In this case for better robustness the singular value decomposition can be used directly for solving  Eq.\eqref{wide equation} \cite{Lawson_Eng}.
 
\section{On the sweep-shape excitation voltage}
\label{About sweep}
The Fourier image of the linear sweep-shape signal is numerically investigated in Ref.  \cite{DFT} (see Fig. 4(e)) and analytically obtained in the general form in Ref.  \cite{rozenberg1982linear}. By setting $r(t)$ in Eq. (9) in Ref.  \cite{rozenberg1982linear}  to rectangle function on sweep-time interval and   by applying  Eqs. 8.250.2-3 from ref.  \cite{Gradshteyn} to result, we obtain difference between Frensel integrals, which is close to rectangle function on $\pm \fb$ interval.

\bibliography{beta}

\end{document}